\documentclass{ws-ijmpa}
\usepackage{graphicx}
\usepackage{epstopdf}
\usepackage{bm}
\usepackage{amsfonts}
\usepackage{lineno,hyperref,hhline}

\begin{document}

\markboth{Ghosh Et Al.}{Gravastars in $f(\mathbb{T},\mathcal{T})$ gravity}

\catchline{}{}{}{}{}

\title{Gravastars in $f(\mathbb{T},\mathcal{T})$ gravity}

\author{Shounak Ghosh}
\address{Department of Physics, Indian Institute of Engineering Science and Technology, Shibpur, B. Garden, Howrah 711103, West Bengal, India\\shounak.rs2015@physics.iiests.ac.in}

\author{A.D. Kanfon}
\address{D\'epartement de Physique, Facult\'e des Sciences et Techniques de Natitingou, BP 72, Natitingou, Benin \\kanfon@yahoo.fr}

\author{Amit Das}
\address{Department of Physics, Government College of Engineering and Ceramic Technology, Kolkata 700010, West Bengal, India               
\\amdphy@gmail.com}

\author{M.J.S. Houndjo}
\address{Facult\'e des Sciences et Techniques de Natitingou, BP 72, Natitingou, Benin, Institut de Math\'ematiques et de Sciences Physiques
(IMSP), Universit\'e  d'Abomey-Calavi Porto-Novo, 01 BP 613 Porto-Novo, Benin
\\sthoundjo@yahoo.fr}

\author{I.G. Salako}
\address{D\'epartement de Physique, Universit\'e Nationaled’ Agriculture, 01 BP 55 Porto-Novo, Benin, Institut de Math\'ematiques et de Sciences Physiques (IMSP), 01 BP 613 Porto-Novo, Benin \\inessalako@gmail.com}

\author{Saibal Ray}
\address{Department of Physics, Government College of Engineering and Ceramic Technology, Kolkata 700010, West Bengal, India                \\saibal@associates.iucaa.in}

\maketitle	

\begin{history}
\received{Day Month Year} \revised{Day Month Year} \comby{Managing Editor}
\end{history}

\begin{abstract}
We propose a stellar model under the $f(\mathbb{T},\mathcal{T})$ gravity following Mazur-Mottola's
conjecture \cite{Mazur2001,Mazur2004} known as gravastar which is generally believed as a viable alternative
to black hole. The gravastar consists of three regions, viz., (I) Interior region, (II) Intermediate shell
region, and (III) Exterior region. The pressure within the interior core region is assumed to be equal to the
constant negative matter-energy density which provides a constant repulsive force over the thin shell region. The shell
is assumed to be made up of fluid of ultrarelativistic plasma and following the  Zel'dovich's conjecture
of stiff fluid \cite{Zeldovich1972} it is also assumed that the pressure which is directly proportional 
to the matter-energy density according to Zel'dovich's conjecture, does
cancel the repulsive force exerted by the interior region. The exterior region is completely vacuum and it can be
described by the Schwarzschild solution. Under all these specifications we find out a set of exact and singularity-free
solutions of the gravastar presenting several physically valid features within the framework of alternative gravity, 
namely $f(\mathbb{T},\mathcal{T})$ gravity ~\cite{Harko2014}, where the part of the gravitational Lagrangian 
in the corresponding action is taken as an arbitrary function of torsion scalar $\mathbb{T}$ and the trace of the energy-momentum tensor $\mathcal{T}$. 
\end{abstract}


\section{Introduction}
It is generally believed that near the end of the stellar evolution if the mass
of the stellar remnant after forming planetary nebula or initiating supernova
explosion exceeds three solar mass, then the star continues to collapse due to
self gravitational pull leading to highly dense object, i.e., black hole. A very
simplest example of it is the Schwarzschild black hole which is assumed to be
static as well as uncharged and can be well defined by the following metric
(for geometrized units $G=c=1$)
\begin{equation}
ds^2=\left(1-\frac{2M}{r}\right)dt^2-\left(1-\frac{2M}{r}\right)^{-1}dr^2 -r^2(d\theta^2+sin^2\theta d\phi^2),\label{eq1}
\end{equation}

where $M$ is the mass of the gravitating object. The metric is singular at $r=0$
(curvature singularity) and $r=2M$ (coordinate singularity).
The surface at $r=2M$ which is known as event horizon, prohibits even light to escape from
it due to massive gravitational pull.

In order to overcome the problem of singularity as well as to avoid the presence of event horizon
Mazur and Mottola~\cite{Mazur2001,Mazur2004} first ever proposed the model of {\it gra}vitatio-nally
{\it va}cuum {\it star} (gravastar) as the alternative to the mysterious end state of gravitationally
collapsing star, i.e., black hole which is generally assumed to overcome any kind of repulsive nonthermal
pressure of degenerate elementary particle, whatsoever. They generated a new kind of solution by extending the
idea of Bose-Einstein condensation and fabricated the model as a cold, dark and compact object with an
interior de Sitter condensate phase as well as Schwarzschild  exterior. Again these two spacetimes were
assumed to be separated by a shell with a small but finite thickness $\ell$ consisting of a stiff
fluid with equation of state (EOS) $p=\rho$ \cite{Zeldovich1972}. It had also been shown that unlike the black hole
the new solution was stable on the basis of thermodynamics and did not possess any information paradox.
The entire structure of gravastar can be envisaged with different EOS for the
different regions of it as follows:\\

\noindent I. Interior ($0 \leq r < r_1$): $p = -\rho$,\\
II. Thin Shell ($r_1\leq r \leq r_2$ ): $p = +\rho$,\\
III. Exterior ($r_2 < r $): $p =\rho= 0$.\\

One can find plenty of works related to gravastar based on different mathematical and physical issues
as available in literature. Most of these works have been done in the framework of Einstein's general
relativity (GR)~\cite{Mazur2001,Mazur2004,Visser2004,Cattoen2005,Carter2005,Bilic2006,Lobo2006,DeBenedictis2006,Lobo2007,Horvat2007,Cecilia2007,Rocha2008,Horvat2008,Nandi2009,Turimov2009,Usmani2011,Lobo2013,Bhar2014,Rahaman2015,Ghosh2017}.
Now-a-days Einstein's general relativity presents some shortcomings in theoretical as well as observational aspects.
In general relativity we often encounter the presence of singularities and there is a lack of self-consistent
theory of quantum gravity. Also, from the observational point of view GR is incapable
to address galactic, extra-galactic and cosmic dynamics without the consideration of the presence of exotic form
of matter-energy which are often known as dark matter and dark energy~\cite{Riess2004,Eisenstein2005,Astier2006,Spergel2007}.
Alternatively, in spite of changing the source side of the Einstein field equations one can equally ask to modify the gravitational sector to describe the galactic as well as cosmic dynamics such as late-time acceleration of the Universe etc.
As a result several modified theories of gravitation such as $f(R)$ gravity, $f(\mathbb{T})$ gravity, $f(R,\mathcal{T})$ gravity
etc. have been proposed from time to time. In all these theories the geometrical part have been changed by taking a generalized
functional form of the argument as the gravitational Lagrangian in the corresponding action. In modified theories of gravity
one usually generalizes the corresponding action on the basis of curvature description of gravity. Interestingly, one can
modify the corresponding action based on torsion, but not on curvature as was first done by Einstein himself and that is known as the ``Teleparallel Equivalent of General Relativity" (TEGR) where the gravity is described by torsion tensor and not by curvature~\cite{Moller1961,Pellegrini1963,Hayashi1979,Arcos2004,Unzicker2005,Maluf2013,Aldrovandi2013,re1,re2,re3,re4}. One can further
modify TEGR by extending $\mathbb{T}$ to an arbitrary function in the Lagrangian~\cite{Ferraro2007,Bengochea2009,Linder2010}.

Moreover, it is possible to modify the general relativity by coupling the geometric sector with the non-geometric sector
~\cite{Uzan1999,Ritis2000,Bertolami2000,Faraoni2000,Amendola1993}. Furthermore, in another modification one can consider
the coupling of the matter Lagrangian to the Ricci scalar~\cite{Bertolami2007,Bertolami2008,Bertolami2010} and extend it to the arbitrary function of $(R,\mathcal{L}_m)$~\cite{Harko2008,Harko2010,Wang2012,Harko2013a,Harko2013b}. Also,
one may consider the models where Ricci scalar $R$ is coupled to the trace of the energy-momentum tensor $\mathcal{T}$ and
further extend it to an arbitrary function such as in $f(R,\mathcal{T})$ theory~\cite{Harko2011}.

Similarly, starting from TEGR but not from GR, it is also possible to consider the matter-coupled
modified gravity theory. One of such theories, namely $f(\mathbb{T},\mathcal{T})$ gravity,
has been first proposed by Harko {\it et al.}~\cite{Harko2014}. In this modified theory the
part of the gravitational Lagrangian is taken as an arbitrary function of torsion scalar
$\mathbb{T}$ and the trace of the energy-momentum tensor $\mathcal{T}$. Comparing with the
other theories based on the formalism of curvature or torsion, $f(\mathbb{T},\mathcal{T})$
seems to be a completely different modification for describing the gravity. In this paper,
we study the gravastar model under $f(\mathbb{T},\mathcal{T})$ gravity, being motivated
by the previous successful work of Das {\it et al.} on compact star~\cite{Das2015,Das2016} as well as
gravastar~\cite{Das2017} under alternative formalism of GR.

Most of the works under the background of $f(\mathbb{T},\mathcal{T})$ gravity
have been done in different cosmological realm as available in literature~\cite{Davood2014,Nassur2015,Salako2015,Ganiou2016a,Ganiou2016b,Junior2016,Diego2016,Farrugia2016,Mirzaei2017}.
Though there are a few astrophysical applications of $f(\mathbb{T},\mathcal{T})$
gravity which can be found in Refs.~\cite{Pace2017a,Pace2017b}. Under the
backgound of $f(\mathbb{T},\mathcal{T})$ gravity Pace and Said~\cite{Pace2017a}
have derived the working model for the Tolman-Oppenheimer-Volkoff (TOV) equation
for quark star incorporating the MIT bag model whereas in another work~\cite{Pace2017b}
they have derived the TOV equation in neutron star systems using a perturbative approach.

The outline of the present investigation is as follows: In Sec. 2 the basic
mathematical formalism of $f(\mathbb{T},\mathcal{T})$ theory has been presented
as the background study. The explicit form of field equations along with the
conservation equation of $f(\mathbb{T},\mathcal{T})$ gravity for the specific
form of gravitational Lagrangian is provided in Sec. 3 whereas in Sec. 4 we
obtain the solutions of the field equations considering the different regions,
viz., interior region, exterior region and the shell region of the gravastar.
In Sec. 5 we discuss the matching conditions in order to determine the numerical
values of several constants which have arisen in different calculations. Sec. 6
deals with some physical features, i.e., proper length, pressure, energy and
entropy within the shell region. . Junction conditions are presented in Sec. 7
and finally, we pass some concluding remarks in Sec. 8.

\section{Basic Mathematical Formalism of $f(\mathbb{T},\mathcal{T})$ Theory}
The modified theories of Tele-Parallel gravity  are those for which the scalar
torsion of Tele-Parallel action is substituted by an arbitrary function of this
latter. As it is done in Tele-Parallel, the modified versions of  this theory are also
described by the orthonormal tetrads and it's components are defined on the tangent
space of each point of the manifold. The line element is written as
\begin{equation}
ds^2=g_{\mu\nu}dx^\mu dx^\nu=\eta_{ij}\theta^i\theta^j,\label{eq2}
\end{equation}
with the following definitions
\begin{equation}
dx^{\mu}=e_{i}^{\;\;\mu}\theta^{i}; \,\quad \theta^{i}=e^{i}_{\;\;\mu}dx^{\mu}.\label{eq3}
\end{equation}

Note that  $\eta_{ij}=diag(1,-1,-1,-1)$ is the Minkowskian metric and the $\{e^{i}_{\;\mu}\}$ are the 
components of the tetrad which satisfy the following identity:
\begin{eqnarray}
e^{\;\;\mu}_{i}e^{i}_{\;\;\nu}=\delta^{\mu}_{\nu},\quad e^{\;\;i}_{\mu}e^{\mu}_{\;\;j}=\delta^{i}_{j}.\label{eq4}
\end{eqnarray}

In GR one use the following Levi-Civita's connection
\begin{equation}
\bar{\Gamma}{}_{\;\;\mu \nu }^{\rho}=\frac{1}{2}g^{\rho \sigma }\left(\partial _{\nu} g_{\sigma \mu}+\partial _{\mu}g_{\sigma \nu}-\partial _{\sigma}g_{\mu \nu}\right),\label{eq5}
\end{equation}
 which  preserves the curvature whereas the torsion vanishes. But in the Tele-Parallel theory and its modified version,
one keeps the scalar torsion by using Weizenbock's connection defined as
\begin{eqnarray}\label{eq6}
\Gamma^{\lambda}_{\mu\nu}=e^{\;\;\lambda}_{i}\partial_{\mu}e^{i}_{\;\;\nu}=-e^{i}_{\;\;\mu}\partial_\nu e_{i}^{\;\;\lambda}.
\end{eqnarray}

From this connection, one obtains the geometric objects. The first is the torsion as defined by
\begin{equation}
T^{\lambda}_{\;\;\;\mu\nu}= \Gamma^{\lambda}_{\mu\nu}-\Gamma^{\lambda}_{\nu\mu},\label{eq7}
\end{equation}
from which we define the second object contorsion defined as
\begin{eqnarray}
K_{\;\;\mu \nu }^{\lambda} \equiv \Gamma _{\;\mu \nu }^{\lambda }
-\bar{\Gamma }{}_{\;\mu \nu }^{\lambda}=\frac{1}{2}(T_{\mu }{}^{\lambda}{}_{\nu }
+ T_{\nu}{}^{\lambda }{}_{\mu }-T_{\;\;\mu \nu }^{\lambda}),\label{eq8}
\end{eqnarray}
where the expression $\bar{\Gamma }{}_{\;\;\mu \nu }^{\lambda}$ designs the above defined connection. Then we can write
\begin{equation}
K^{\mu\nu}_{\;\;\;\;\lambda}=-\frac{1}{2}\left(T^{\mu\nu}_{\;\;\;\lambda}-T^{\nu\mu}_{\;\;\;\;\lambda}+T^{\;\;\;\nu\mu}_{\lambda}\right).\label{eq9}
\end{equation}

The two previous geometric objects (the torsion and the contorsion) are used to define another tensor by
\begin{equation}
S_{\lambda}^{\;\;\mu\nu}=\frac{1}{2}\left(K^{\mu\nu}_{\;\;\;\;\lambda}+
\delta^{\mu}_{\lambda}T^{\alpha\nu}_{\;\;\;\;\alpha}-\delta^{\nu}_{\lambda}T^{\alpha\mu}_{\;\;\;\;\alpha}\right).\label{eq10}
\end{equation}

The torsion scalar is usually constructed from torsion and contorsion as follows:
\begin{equation}
\mathbb{T} = S_\sigma^{~\mu\nu} T^\sigma_{~\mu\nu}.\label{eq11}
\end{equation}

In the modified versions of Tele-Parallel gravity, one can use a general algebraic
function of scalar torsion instead of the scalar torsion only as it is done in the initial theory.
So, the modified action~\cite{Harko2014} can be written as
\begin{equation}
\mathbb{S}=\int d^4x e\left[\frac{\mathbb{T}+f(\mathbb{T},\mathcal{T})}{16\pi}+\mathcal{L}_{m} \right].\label{eq12}
\end{equation}

Varying the action with respect to the tetrad, one obtains the equations of motion~\cite{Harko2014} as
\begin{eqnarray}\label{eq13}
&&[\partial_\xi(ee^\rho_a S^{\;\;\sigma\xi}_\rho)-ee^\lambda_a S^{\rho\xi\sigma} T_{\rho\xi\lambda}](1+f_{\mathbb{T}})
+ e e^\rho_a(\partial_\xi\mathbb{T})S^{\;\;\sigma\xi}_\rho f_{\mathbb{T}\mathbb{T}} +\frac{1}{4} e e^\sigma_a (\mathbb{T}) =\nonumber\\
&&- \frac{1}{4} e e^\sigma_a f -e e^\rho_a(\partial_\xi \mathcal{T})S^{\;\;\sigma\xi}_\rho f_{\mathbb{T}\mathcal{T}} +f_{\mathcal{T}}\;\left(\frac{e\,T^\sigma_{\;\;a}  + e e^\sigma_a \;p }{2}\right) + 4\pi\,e\,T^\sigma_{\;\;a} \;,
\end{eqnarray}
with
$f_{\mathcal{T}} = \partial f/\partial \mathcal{T} $,
$f_{\mathbb{T}} = \partial f/\partial\mathbb{T}$,
$ f_{\mathbb{T}\mathcal{T}} = \partial^{2}f/\partial\mathbb{T}\partial \mathcal{T}$,
$f_{\mathbb{T}\mathbb{T}}  = \partial^{2}f/\partial \mathbb{T}^{2}$
and $T^\sigma_{\;\;a}$ is the energy-momentum tensor of
matter field where we assume the energy-momentum tensor
to be that of a perfect fluid, i.e.,
\begin{equation}\label{eq14}
T_{\mu\nu}=(\rho+p)u_\mu u_\nu-pg_{\mu\nu}.
\end{equation}

After some contraction, we can rewrite the first term on the R.H.S. of  Eq. (13) as follows~\cite{Ganiou16}:
\begin{eqnarray}\label{eq15}
 e^a_\nu e^{-1}\partial_\xi(ee^\rho_a
S^{\;\;\sigma\xi}_\rho)-S^{\rho\xi\sigma}T_{\rho\xi\nu} = -\nabla^\xi S_{\nu\xi}^{\;\;\;\;\sigma}-S^{\xi\rho\sigma}K_{\rho\xi\nu},
\end{eqnarray} 

and one can have

\begin{eqnarray}\label{eq16}
G_{\mu\nu}-\frac{1}{2}\,g_{\mu\nu}\,\mathbb{T}
=-\nabla^\rho S_{\nu\rho\mu}-S^{\sigma\rho}_{\;\;\;\;\mu}K_{\rho\sigma\nu}.
\end{eqnarray}

Hence from the combination of Eqs.~(\ref{eq15}) and
(\ref{eq16}), the field equation of Eq. (\ref{eq13}) can be
written as
\begin{equation}
A_{\mu\nu}(1+ f_\mathbb{T}) +\frac{1}{4}g_{\mu\nu}\;\mathbb{T} =B_{\mu\nu}^{eff},\label{eq17}
\end{equation}
where
\begin{eqnarray}\label{eq18}
A_{\mu \nu }=g_{\sigma\mu}e^a_\nu[e^{-1}\partial_\xi(ee^\rho_a
S^{\;\;\sigma\xi}_\rho)-e^\lambda_a S^{\rho\xi\sigma} T_{\rho\xi\lambda}] =-\nabla^\sigma S_{\nu\sigma\mu }-S_{\;\;\;\;\mu }^{\rho\lambda }K_{\lambda \rho \nu }=G_{\mu \nu }-\frac{1}{2}g_{\mu \nu}\mathbb{T}
\end{eqnarray}

and

\begin{eqnarray}\label{eq19}
B_{\mu\nu}^{eff} =  S^{\rho}_{\;\;\;\mu\nu}\; f_{\mathbb{T}\mathcal{T}}\; \partial_{\rho} \mathcal{T}  -
S^{\rho}_{\;\;\;\mu\nu}\;f_{\mathbb{T}\mathbb{T}}\; \partial_{\rho}\mathbb{T}
- \frac{1}{4} g_{\mu \nu }f + f_{\mathcal{T}}\;\Big(\frac{T_{\mu \nu } + g_{\mu \nu } \;p }{2}\Big) + 4\pi T_{\mu \nu }.
\end{eqnarray}

So Eq. (\ref{eq17}) takes the following form
\begin{equation}
(1+ f_{\mathbb{T}})\,G_{\mu\nu}=T_{\mu\nu}^{eff},\label{eq20}
\end{equation}
where
\begin{eqnarray}\label{eq21}
T_{\mu\nu}^{eff} =S^{\rho}_{\;\;\;\mu\nu}\; f_{\mathbb{T}\mathcal{T}}\; \partial_{\rho} \mathcal{T}  -
S^{\rho}_{\;\;\;\mu\nu}\;f_{\mathbb{T}\mathbb{T}}\; \partial_{\rho}\mathbb{T}+\frac{1}{4} g_{\mu \nu } \Big(\mathbb{T}-f \Big)  +\frac{\mathbb{T}\,g_{\mu\nu}\,f_{\mathbb{T}}}{2}  \nonumber \\
+f_{\mathcal{T}}\;\Big(\frac{T_{\mu \nu } + g_{\mu \nu } \;p }{2}\Big)+ 4\pi T_{\mu \nu } \;.
\end{eqnarray}

The covariant derivative of Eq. (\ref{eq20}) reads as
\begin{eqnarray}\label{eq22}
&&\nabla^\mu \Big[(1+ f_{\mathbb{T}})\, G_{\mu\nu}\Big] =  \nabla^\mu T_{\mu\nu}^{eff}= \nabla^\mu  \Bigg[\frac{\mathbb{T}\,g_{\mu\nu}\,f_{\mathbb{T}}}{2}+ S^{\rho}_{\;\;\;\mu\nu}\; f_{\mathbb{T}\mathcal{T}}\; \partial_{\rho} \mathcal{T}- S^{\rho}_{\;\;\;\mu\nu}\;f_{\mathbb{T}\mathbb{T}}\; \partial_{\rho}\mathbb{T} \nonumber\\
&+& \frac{1}{4} g_{\mu \nu } \Big(\mathbb{T}-f \Big) + f_{\mathcal{T}}\;\Big(\frac{T{\mu \nu }+ g_{\mu \nu } \;p }{2}\Big) + 4\pi T_{\mu \nu }\Big].
\end{eqnarray}

The previous equation leads to the following expression
\begin{eqnarray}\label{eq23}
\nabla^\mu T_{\mu \nu } &=& \frac{-2}{(f_{\mathcal{T}}+8\pi) }
\Bigg\{ \nabla^\mu  \Big[\frac{\mathbb{T}\,g_{\mu\nu}\,f_{\mathbb{T}}}{2} + S^{\rho}_{\;\;\;\mu\nu}\; f_{\mathbb{T}\mathcal{T}}\;
\partial_{\rho} \mathcal{T}- S^{\rho}_{\;\;\;\mu\nu}\;f_{\mathbb{T}\mathbb{T}}\; \partial_{\rho}\mathbb{T}+\frac{1}{4} g_{\mu \nu } \Big(\mathbb{T}- f \Big)\Big]   \nonumber\\
&+&\Big(\frac{T_{\mu\nu }+ g_{\mu\nu } \;p }{2}\Big)\,\nabla^\mu f_{\mathcal{T}} + \frac{f_{\mathcal{T}}}{2}\,\nabla^\mu ( g_{\mu \nu } \;p )- G_{\mu\nu} \nabla^\mu (1+ f_{\mathbb{T}})\Bigg\}.
\end{eqnarray}

We have taken the functional form of $f(\mathbb{T},\mathcal{T})$ as $f(\mathbb{T},\mathcal{T})= g_1(\mathbb{T})+ g_2(\mathcal{T})$ where for simplicity we have chosen $g_1(\mathbb{T})=0$ and $g_2(\mathcal{T})= 2Q\mathcal{T}$ so that $f_{\mathbb{T}}=f_{\mathbb{T}\mathbb{T}} =f_{\mathbb{T}\mathcal{T}}= 0$.

For the above form of $f(\mathbb{T},\mathcal{T})$ the Eq. (\ref{eq23}) reduces to
\begin{eqnarray}\label{eq24}
\nabla^\mu T^\mu_\nu = \frac{1}{(f_{\mathcal{T}}+ 8\pi) } \Bigg\{
\frac{1}{2} \delta^\mu_\nu \nabla^\mu  f(\mathbb{T},\mathcal{T})&-&\Big(T^\mu_\nu+ \delta^\mu_\nu\;p \Big)\,
\nabla^\mu f_{\mathcal{T}} - f_{\mathcal{T}}\,\delta^\mu_\nu  \nabla^\mu \,p \Bigg\}.
\end{eqnarray}

\section{The field equations in $f(\mathbb{T},\mathcal{T})$ gravity}

Assuming that the manifold to be static and spherically symmetric, the metric can be written as
\begin{equation}
ds^{2}=e^{\nu(r)}dt^{2}-e^{\lambda(r)}dr^{2}-r^{2}\left(d\theta^{2}+\sin^{2}\theta d\phi^{2}\right).\label{eq25}
\end{equation}

In order to re-write the line element Eq. (\ref{eq25}) into the invariant form under the Lorentz transformations as in Eq. (\ref{eq2}),
we define the tetrad matrix $[e^{i}_{\;\mu}]$ as
\begin{equation}
\left[e^{i}_{\;\;\mu}\right]= diag \left[e^{\nu(r)/2},e^{\lambda(r)/2},r,r\sin\theta\right].\label{eq26}
\end{equation}

From  Eq. (\ref{eq26}), one can obtain $e=\det{\left[e^{i}_{\;\;\mu}\right]}=e^{(\nu+\lambda)/2}r^2 \sin\theta$ and using Eq. (\ref{eq11}) the  torsion scalar can be written as
\begin{equation}
\mathbb{T}(r)= \frac{2e^{-\lambda}}{r}\left(\nu^{\prime}+\frac{1}{r}\right).\label{eq27}
\end{equation}

Now the nonzero components of the Einstein tensors can be written as

\begin{equation}
G_0^{0}=\frac{e^{-\lambda}}{r^{2}}(-1+e^{\lambda}+\lambda'r),\label{eq28}
\end{equation}

\begin{equation}
G_1^{1}=\frac{e^{-\lambda}}{r^{2}}(-1+e^{\lambda}-\nu'r),\label{eq29}
\end{equation}

\begin{equation}
G_2^{2}=G_3^{3}=\frac{e^{-\lambda}}{4r}[2(\lambda'-\nu')-(2\nu''+\nu'^{2}-\nu'\lambda')r],\label{eq30}
\end{equation}
where primes stand for derivative with respect to the radial co-ordinate $r$.

For the functional form of $f(\mathbb{T},\mathcal{T})$ as
$f(\mathbb{T},\mathcal{T})= g_1(\mathbb{T})+ g_2(\mathcal{T})$
with $g_1(\mathbb{T})=0$ the field Eq. (\ref{eq20}) takes the form
\begin{eqnarray}
(1+ 0)G_{00} &=&T_{00}^{eff}, \label{eq31} \\
(1+ 0)G_{11} &=&T_{11}^{eff}, \label{eq32}\\
(1+ 0)G_{22} &=&T_{22}^{eff},\label{eq33} \\
(1+ 0)G_{33}&=&T_{33}^{eff},\label{eq34}
\end{eqnarray}
where
\begin{equation}
T_0^{0~eff}=\frac{1}{4}\left(\mathbb{T}-f\right)+\left(4\pi+\frac{f_{\mathcal{T}}}{2}\right)\rho+\frac{pf_{\mathcal{T}}}{2},
\end{equation}\label{eq35}
\begin{equation}
T_{1}^{1~eff}=T_{2}^{2~eff}=T_{3}^{3~eff}=\frac{1}{4}\left(\mathbb{T}-f\right)-4\pi\,p. \label{eq36}
\end{equation}

Finally, we have the following field equations:
\begin{eqnarray}
&&-\frac{3e^{-\lambda}}{2r^2}+\frac{1}{r^2}+\frac{e^{-\lambda}}{r}\left(\lambda'-\frac{\nu'}{2}\right)=\rho\left(4\pi+\frac{Q}{2}\right)+\frac{5Qp}{2},\label{eq37}\\
&&-\frac{3e^{-\lambda}}{2r^2}+\frac{1}{r^2}-\frac{3\nu'e^{-\lambda}}{2r}=p\left(\frac{3Q}{2}-4\pi\right)-\frac{Q\rho}{2},\label{eq38} \\
&&\frac{\lambda'e^{-\lambda}}{2r}-\frac{3\nu'e^{-\lambda}}{2r}-\frac{e^{-\lambda}}{4}\left(2\nu'' +\nu'^2-\nu'
\lambda'\right)-\frac{e^{-\lambda}}{2r^2}=p\left(\frac{3Q}{2}-4\pi\right)-\frac{Q\rho}{2}.\label{eq39}
\end{eqnarray}

Also the conservation equation in $f(T,\mathcal{T})$ gravity takes the form as follows
\begin{equation}
\frac{1}{2}(p+\rho)\nu'+p'=\frac{Q}{(8\pi+2Q)}(5p'-\rho') .\label{eq40}
\end{equation}

\section{The solutions of the field equations for different regions of the gravastar}
\subsection{Interior Space-time}
Following the prescription of Mazur and Mottola \cite{Mazur2001,Mazur2004}
we have the EOS that represent the interior
region of the gravastar is given by
\begin{equation}
p=-\rho.\label{eq41}
\end{equation}

Now using Eqs. (\ref{eq40}) and (\ref{eq41}) we have obtained
the matter density and pressure within the shell as
\begin{equation}
p=-\rho=-\rho_c ,\label{eq42}
\end{equation}
where $\rho_c$ is a constant, i.e., the matter density as well as
pressure inside the shell remain constant. This indicates that the
pressure and the density is homogeneous and isotropic inside the gravastar.

Now using Eqs. (\ref{eq37}) and (\ref{eq38}) and putting
the condition provided in Eq. (\ref{eq41}) we can find the metric function
$e^{-\lambda}$ for the interior spacetime of the gravastar as
\begin{equation}
e^{-\lambda}=\frac{2}{3}-\frac{4\rho_c (2\pi-Q)r^2}{9}+\frac{c_1}{r},\label{eq43}
\end{equation}
where $c_1$ is a constant of integration. To make the solution
regular at the centre of the gravastar, i.e., at $r=0$ we have to
set $c_1$ to zero so that
\begin{equation}
e^{-\lambda}=\frac{2}{3}-\frac{4\rho_c (2\pi-Q)r^2}{9}.\label{eq44}
\end{equation}

\begin{figure}[thbp]
\centering
\begin{minipage}[b]{0.6\textwidth}
    \includegraphics[width=\textwidth]{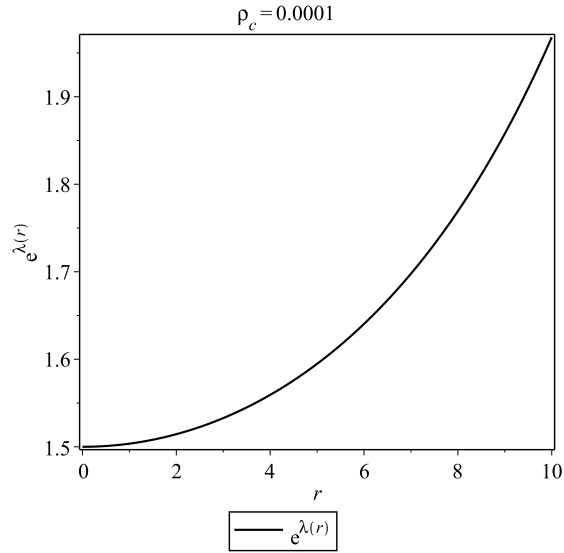}
    \caption{Variation of the metric function ($e^{\lambda(r)}$) within the interior region with respect to $r$ (km).}\label{fig1}
  \end{minipage}
\end{figure}

Now putting the above expression in Eq. (\ref{eq38}) we get
\begin{equation}
e^{\nu}=c_2\left[2 \rho_c(2\pi -Q)r^2-3\right]^{\frac{3\pi-2Q}{2\rho_c(2\pi-Q)}},\label{eq45}
\end{equation}
where $c_2$ is an integration constant. From the above solutions as written by the
Eqs. (\ref{eq44}) and (\ref{eq45}) one can observe that both solutions are regular
at the center (i.e., $r=0$) of the gravastar which means solutions have no singularity at the center.

Now one can calculate the active gravitational mass of the interior of the gravastar as
\begin{equation}
\tilde{M}=\int_0^{R} 4\pi r^2 \rho dr=\frac{4}{3}\pi R^3 \rho_c,\label{eq46}\\
\end{equation}
where $r=R$ is the internal radius of the gravastar.

\subsection{Intermediate thin shell}
The shell has been formed within the intermediate region of two
spacetimes, namely, the interior and the exterior spacetime. The
thickness of the shell is very small but finite and it contains
the entire matter of the collapsing star. In connection to the cold,
compact baryonic universe, this type of ultra relativistic matter can be
termed as the stiff fluid which was first introduced by Zel'dovich~\cite{Zeldovich1972}.
This type of fluid obeys the EOS $p=\rho$ which means it provides the
necessary balancing force to counter the repulsion from the interior
keeping the system in stable equilibrium.

In the present scenario we can set
up an argument that it may happen from thermal excitations at
$0$ K temperature with negligible chemical potential or from the
conserved number density of the gravitational quanta~\cite{Mazur2001,Mazur2004}.
In many cosmological and astrophysical studies various researchers
have used this type of fluid~\cite{Wesson1978,Braje2002,Linares2004} for
feasible explanations.

 Now we are looking for the mathematical as well as physical
 acceptable solution to explain the formation of the shell.
 It is very difficult to obtain the exact solution of the
 field equations within the nonvacuum region using the EOS
 $p=\rho$. However, considering the framework of thin shell
 approximation, we can find the analytical solution of the
 shell. Therefore, we have considered  $e^{-\lambda}$ lying
 between the range $0< e^{-\lambda}\ll1$. Following the prescription
 of Israel~\cite{Israel1966} we can have an argument
 regarding the formation of shell at the intermediate region of the
 two spacetimes (here the vacuum interior and Schwarzschild exterior).

Now with the thin shell approximation, the field Eqs. (\ref{eq37})-(\ref{eq39}) can be written as
\begin{eqnarray}
\frac{1}{r^2}+\frac{e^{-\lambda}\lambda'}{r}=\rho\left(4\pi+\frac{Q}{2}\right)+\frac{5Qp}{2},\\\label{eq47}
\frac{1}{r^2}=p\left(\frac{3Q}{2}-4\pi\right)-\frac{Q\rho}{2},\\\label{eq48}
\frac{e^{-\lambda}\lambda'}{2r}+\frac{e^{-\lambda} \nu' \lambda'}{4}=p\left(\frac{3Q}{2}-4\pi\right)-\frac{Q\rho}{2}.\label{eq49}
\end{eqnarray}

From Eq. (\ref{eq47}) we can calculate the metric function $e^{-\lambda}$ as
 \begin{equation}
e^{-\lambda}=k_1+2\left(\frac{4\pi +Q}{4\pi -Q}\right) \ln r ,\label{eq50}
 \end{equation}
where $k_1$ is a constant of integration which can be calculated using
the matching condition between the thin shell region and the exterior vacuum region.

\begin{figure}[thbp]
\centering
\begin{minipage}[b]{0.6\textwidth}
    \includegraphics[width=\textwidth]{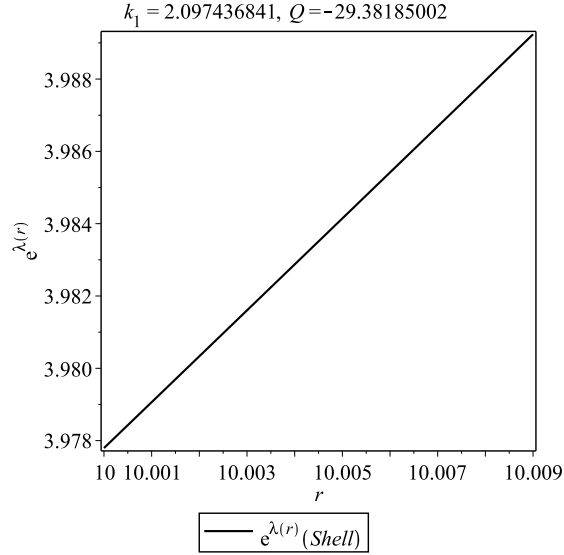}
    \caption{Variation of $e^{\lambda(r)}$ within the shell with respect to $r$ (km).}\label{fig2}
  \end{minipage}
\end{figure}

Now using Eqs. (\ref{eq49}) and (\ref{eq50}) we have found the other metric potential $e^{\nu}$ as
 \begin{equation}
e^{\nu}=k_2r^{-\frac{16\pi}{Q+4\pi}},\label{eq51}
 \end{equation}
where $k_2$ is also a constant of integration. Variation of $e^{\nu(r)}$ has been studied and found to be similar in its nature as Fig. \ref{fig2}.

\subsection{Exterior region}
The exterior of the gravastar is assumed to obey the EOS
$p=\rho=0$ which means that the outside region of the shell is completely vacuum. The
corresponding line-element can be written as
\begin{eqnarray}
ds^2=\left(1-\frac{2M}{r}\right)dt^2&-&\left(1-\frac{2M}{r}\right)^{-1}dr^2-r^2(d\theta^2+\sin^2\theta d\phi^2),\label{eq52}
\end{eqnarray}
which is same as in Schwarzschild type  vacuum solution in $(3+1)$
dimension. Here $M$ is the total mass of the gravastar.

\section{Matching Condition}
To determine the value of the integration constant $k_1$ we have matched the
metric potential at the junction of interior region and the shell. We have obtained
\begin{equation}
k_1=1-\frac{2M}{R_2}+2\left(\frac{Q+4\pi}{Q-4\pi}\right)\ln R_2. \label{eq53}
\end{equation}

We have calculated the value of the other integration constant by matching
the metric potentials at the junction of the shell and the exterior region of the gravastar as
\begin{equation}
k_2=\left(1-\frac{2M}{R_2}\right)R_2^{\frac{16\pi}{4\pi+Q}}. \label{eq54}
\end{equation}

The expression of $Q$, using the continuity condition of $\frac {\delta g_{tt}}{\delta r}$
at the junction of the shell, can be found as
\begin{equation}
 Q=\frac{1}{2} \frac{R_2 \pi(-16+\frac {32M}{R_2})} {M} -4\pi. \label{eq55}
\end{equation}

In the present paper we have used the total mass of the gravastar
$M=3.75M_\odot$ ($M_\odot$ is the solar mass), internal radius $R_1=10$ km
and the thickness of the shell, i.e., $\epsilon=R_2-R_1=0.009$ km and $\rho_c=0.0001$ \cite{Ghosh2018}. Using these values we have determined $k_1$, $k_2$, and $Q$ as $2.097436841$, $0.2562733037 \times 10^{-3}$ and $ -29.38185002 $ respectively.

\section{Physical features of the shell}

\subsection{Proper Thickness}
According to the conjecture of Mazur and Mottola~\cite{Mazur2001,Mazur2004},
the stiff fluid of the shell is situated between the junction of two spacetimes.
The length span of the shell is from $r=R$ (i.e., the phase boundary between
the interior and the shell) to $r=R+\epsilon$ (i.e., the phase boundary
between the shell and the exterior spacetime). So, one can calculate the proper
thickness between these two interfaces and proper length or the proper thickness
of the shell can be determined using the following formula
\begin{eqnarray}
 \ell &=&\int^{R+\epsilon}_R \frac{dr}{\sqrt{e^{-\lambda}}} 
      =\int^{R+\epsilon}_R \frac{dr}{\sqrt{k_1+2\left(\frac{4\pi +Q}{4\pi -Q}\right) \log r}}  \nonumber\\
      &=&\left[\frac{e^{-\frac{k_1}{b}}erfc\left[-\sqrt{\frac{k_1+b\log r}{b}}\right]\sqrt{-\frac{k_1+b\log r}{b}}}{\sqrt{k_1+b\log r }}\right]_R^{R+\epsilon},\label{eq56}
\end{eqnarray}
where $b=2\left(\frac{4\pi +Q}{4\pi -Q}\right)$ and $erfc$ is the complementary error function which can be defined for a variable $z$ as
 \begin{equation}
erfc(z)=\frac{\Gamma(\frac{1}{2},z^2)}{\sqrt{\pi}}. \label{eq57}
 \end{equation}

It has been observed that the proper length within the shell remains positive and finite whose variation found to be similar to that of the matter density.

\subsection{Pressure and density}
\begin{figure}[thbp]
  \centering
  \begin{minipage}[b]{0.6\textwidth}
    \includegraphics[width=\textwidth]{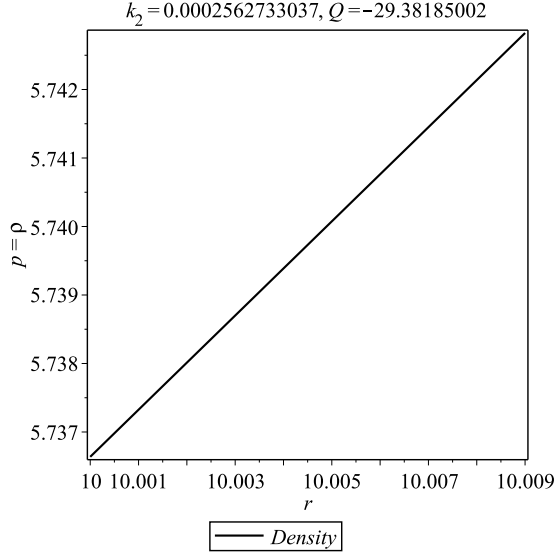}
    \caption{Variation of the matter density ($\rho$) of the shell with respect to $r$ (km). }\label{fig5}
  \end{minipage}
\end{figure}

From Eqs. (\ref{eq40}) and (\ref{eq51}) we obtain the matter density as well as the pressure of the shell as
 \begin{equation}
\rho=p=k_3(e^{-\nu})^{\frac{4\pi+Q}{4\pi-Q}}=k_3 k_2^{\frac{Q+4\pi}{Q-4\pi}}r^{\frac{16\pi}{4\pi-Q}}=\rho_0 r^{\frac{16\pi}{4\pi-Q}}, \label{eq58}
 \end{equation}
where $k_3$ is an integration constant and $\rho_0=k_3 k_2^{(\frac{Q+4\pi}{Q-4\pi})}$ is also constant. From the above
equation one can see that the matter density increases with $r$ over the shell.
Eq. (\ref{eq58}) suggests that the density of the shell must be very high. The
variation of the density in Fig. \ref{fig5} also indicates that it increases from the
interior boundary to the exterior boundary, i.e., the shell is more compact at
the junction of exterior region than that of the interior region.

\subsection{Energy}
We calculate the energy content within the thin shell as
\begin{eqnarray}
E= \int^{R+\epsilon}_R 4\pi r^2 \rho dr = \int^{R+\epsilon}_R 4\pi r^2 \rho_0 r^{\frac{16\pi}{Q+4\pi}} dr = \frac{4 \pi \rho_0 (4 \pi-Q)}{28 \pi-3Q}\left[r^\frac{28 \pi-3Q}{4 \pi-Q}\right]_R^{R+\epsilon}. \label{eq59}
\end{eqnarray}

It has been verified that variation of the energy with respect to the radial parameter $r$ shows similar nature as the matter density within the shell.

\subsection{Entropy}
The entropy within the thin shell can be obtained using the following equation as
\begin{equation}
 S=\int^{R+\epsilon}_R 4\pi r^2 \frac{s(r)}{\sqrt{e^{-\lambda}}} dr,\label{eq60}
\end{equation}
where $s(r)$ is the entropy density.

So, following Mazur and Mottola~\cite{Mazur2001,Mazur2004} we can write it as follows
\begin{equation}
 s(r)=\frac{\xi^2 k_B^2 T(r)}{4\pi\hbar^2}=\xi\frac{k_B}{\hbar}\sqrt{\frac{p}{2\pi}},\label{eq61}
\end{equation}
where $\xi$ is a dimensionless constant.

\begin{figure}[!tbp]
  \centering
  \begin{minipage}[b]{0.6\textwidth}
    \includegraphics[width=\textwidth]{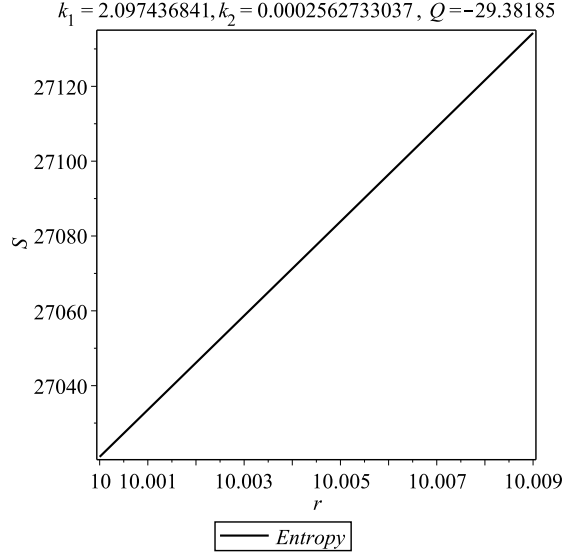}
    \caption{Variation of the entropy ($S$) of the shell with respect to $r$ (km).}\label{fig7}
  \end{minipage}
\end{figure}

Inserting Eq. (\ref{eq61}) in Eq. (\ref{eq60}) one can calculate the entropy of the fluid within the shell as
\begin{eqnarray}
S&=&\frac{4\pi \xi k_B}{\hbar \sqrt{2 \pi}}\int^{R+\epsilon}_R  r^2 \sqrt{\frac{p}{ e^{-\lambda}}} dr =\frac{\xi k_B \sqrt{8\pi}}{\hbar}\int^{R+\epsilon}_R r^2\sqrt{\frac{\rho_0 r^{\frac{16\pi}{4\pi-Q}}}
{k_1+2\left(\frac{4\pi+Q}{4\pi-Q}\right)\ln r}}dr \nonumber\\
&=&\frac{\xi k_B \sqrt{8\pi \rho_0}}{\hbar b\sqrt{-\frac{(6+c)(k_1+\log r)}{b}}}\left[e^{-\frac{k_1(c+b)}
{2b}}\sqrt{2\pi(k_1+\log r)}  \right. \left. erfc \left[-\frac{(3+\frac{c}{2})(k_1+b\log r)}{b}\right] \right]_R^{R+\epsilon},\label{eq62}
\end{eqnarray}
where $c={16 \pi}/(4\pi-Q)$. In Planckian units we have $k_B=\hbar=1$.
The variation of the entropy over the thin shell has been shown in Fig. \ref{fig7}.

\section{Junction Condition}
We have three segments of the gravastar, the shell of the gravastar is
formed between the junction of two spacetimes, i.e., the interior and
the exterior regions. Using the condition of  Darmois-Israel~\cite{Darmois1927,Israel1966}
we calculate the surface stresses at the junction interface. The
intrinsic surface stress energy tensor $S_{ij}$ is given by the Lanczos
equation~\cite{Lanczos1924,Sen1924,Perry1992,Lake1996} in the following form
\begin{equation}
S_{j}^{i}=-\frac{1}{8\pi} (K_{j}^{i}-\delta_{j}^{i} K_{k}^{k}),\label{eq63}
\end{equation}
the discontinuity in the second fundamental form is given by
\begin{equation}
K_{ij}=K_{ij}^{+}-K_{ij}^{-},\label{eq64}
\end{equation}
where the second fundamental form is given by
\begin{equation}
K_{ij}^{\pm}=-n_{\nu}^{\pm}\left[\frac{\partial^{2}X_{\nu}}{\partial
\xi^{i}\partial\xi^{j}}+\Gamma_{\alpha\beta}^{\nu}\frac{\partial
X^{\alpha}}{\partial \xi^{i}}\frac{\partial X^{\beta}}{\partial
\xi^{j}} \right]|_S,\label{eq65}
\end{equation}
where the unit normal vector $n_{\nu}^{\pm}$ are defined as
\begin{equation}
n_{\nu}^{\pm}=\pm\left|g^{\alpha\beta}\frac{\partial f}{\partial
X^{\alpha}}\frac{\partial f}{\partial X^{\beta}}
\right|^{-\frac{1}{2}}\frac{\partial f}{\partial X^{\nu}},\label{66}
\end{equation}
with $n^{\nu}n_{\nu}=1$. Where $\xi^{i}$ represent the intrinsic
coordinates on the shell and the parametric equation of the shell
is$f(x^{\alpha}(\xi^{i}))=0$ coordinate on the shell. Here `$+$'
and `$-$' stands for the exterior Schwarzschild spacetime and the
interior de Sitter spacetime of the gravastar respectively.

Now using Lanczos equation \cite{Lanczos1924} for a spherically
symmetric spacetime, the surface stress energy tensor can be written
as $S_{i}^{j}= diag (\Sigma,-\textit{P},-\textit{P},-\textit{P})$.
Here $\Sigma$ represents the surface energy density  and $\textit{P}$
stands for the surface pressure. So, we can express $\Sigma$ and $\textit{P}$
by the following equations:
\begin{equation}
\Sigma=\left[ -\frac{1}{4\pi R}\sqrt{f}\right]^+_-, \label{eq67}
\end{equation}
and
\begin{equation}
\textit{P}=\left[-\frac{\Sigma}{2}+\frac{f^{'}}{8
\pi\sqrt{f}}\right]^+_-. \label{eq68}
\end{equation}

So, using the above two equations we have eventually obtained the surface energy
density and surface pressure respectively as
\begin{equation}
\Sigma=-\frac{1}{4\pi R}\left[\sqrt{1-\frac{2M}{R}}-\sqrt{\frac{2}{3}-\frac{4\rho_c (2\pi-Q)r^2}{9}}\right],\label{eq69}
\end{equation}
and
\begin{equation}
\textit{P}=\frac{1}{8\pi R}\left[\frac{1-\frac{M}{R}}{ \sqrt{1-\frac{2M}{R}}}- \frac{\frac{2}{3}-\frac{8\rho_c(2\pi-Q)R^2}{9}}{\sqrt{\frac{2}{3}-\frac{4\rho_c(2\pi-Q)R^2}{9}}}\right].\label{eq70}
\end{equation}

\begin{figure}[!tbp]
  \centering
  \begin{minipage}[b]{0.6\textwidth}
    \includegraphics[width=\textwidth]{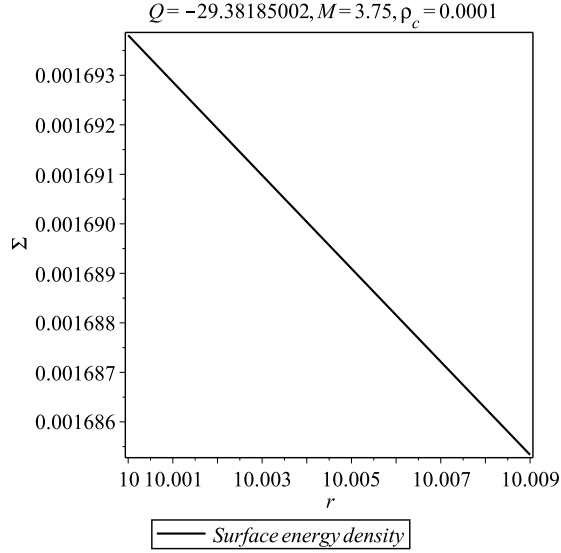}
    \caption{Variation of the surface energy density ($\Sigma$) of the shell with respect to $r$ (km).}\label{fig8}
  \end{minipage}
\end{figure}

Variation of the surface energy density is shown in Fig. 4. We also note that variation
of the  surface pressure is exactly similar in nature as the surface energy density. There is a discontinuity of the second fundamental
form at the junction between the two spacetimes which further implies
that there must be a matter component (ultra relativistic fluid) obeying
EOS $p=\rho$. This noninteracting matter or the fluid characterizes the
existence of the thin shell of the gravastar.

Now one can calculate the mass of the thin shell using the surface
energy density of Eq. (\ref{eq69}) as
\begin{equation}
m_s=4\pi r^2 \Sigma =R\left[\sqrt{\frac{2}{3}-\frac{4\rho_c (2\pi-Q)R^2}{9}}-\sqrt{1-\frac{2M}{R}}\right].\label{eq71}
\end{equation}

With the help of the above equation we can also calculate the total mass
of the gravastar in terms of the mass of the thin shell as
\begin{equation}
M=\frac{R}{6}-\frac{m_s^2}{2R}+m_s\sqrt{\frac{2}{3}-\frac{4\rho_cR^2(2\pi-Q)}{9}}+\frac{2 \rho_c (2\pi-Q)R^3}{9}.\label{eq72}
\end{equation}

\section{Conclusions}
In the present work we provide a unique stellar model following the conjecture
of Mazur-Mottola~\cite{Mazur2001,Mazur2004} under the $f(\mathbb{T},\mathcal{T})$
gravity. The model termed as gravastar by them is assumed to consists of three
distinct regions, namely, (i) interior core region, (ii) intermediate shell
region, and (iii) exterior vacuum region and each region is governed by specific
EOS. With all these specifications we find out a set of exact and singularity-free
solutions presenting several interesting and valid properties of the gravastar
within in the framework of $f(\mathbb{T},\mathcal{T})$ gravity.

We have noted down several salient aspects of the solution set studying the
above mentioned structural form of a gravastar and those can be described below:\\

\indent (1) {\it Interior region}: Using the EOS of (\ref{eq41}) along with the
conservation equation of Eq. (\ref{eq40}) we have found that the matter density
as well as the pressure remains constant in the interior. Again using
Eqs. (\ref{eq37})- (\ref{eq41}) we obtain the metric functions
$e^{\lambda(r)}$ and $e^{\nu(r)}$. From these equations it is clear that both the functions
are continuous at the origin $r=0$, i.e., free from any central singularity.
The variation of the metric function $g_{rr}$ has been shown in Fig. \ref{fig1}.

\indent (2) {\it Intermediate thin shell}: Using the thin shell approximation we
have solved Einstein's field equations to obtain the metric functions. The variation
of the metric functions $e^{\lambda(r)}$ is shown in Fig. \ref{fig2}. Both the parameters remain finite and
positive over the shell. The results provide a physically acceptable solution
for the formation of gravastar under $f(\mathbb{T},\mathcal{T})$ gravity.

\indent (3) {\it Proper thickness}:The proper thickness or proper length $\ell$ of the shell 
is found to be gradual increasing in nature from the interior junction to the exterior junction.

\indent (4) {\it Pressure-density}: The pressure and density ($p=\rho=0$) of the
ultrarelativistic fluid in the shell is plotted with respect to the radial
coordinate $r$ in Fig. \ref{fig5} which shows that the matter density remains positive
and increases gradually with the thickness of the shell. This suggests that
the shell becomes more denser at the exterior boundary than the interior boundary.

\indent (5) {\it Energy}: The energy of the shell is proportional to the radial
coordinate $r$ which demands that the energy must be higher at the exterior boundary.
The variation of energy is similar as the matter density shown in Fig. 3 and satisfies the requirement that the energy of the shell
increases with the radial parameter of the shell.

\indent (6) {\it Entropy}: The entropy $S$ within the shell has been obtained in Eq. (\ref{eq60}) and 
the variation is found similar as Fig. 4 which shows that the entropy is gradually
increasing with respect to the radial coordinate $r$ indicating a maximum value on the surface
of the gravastar that fulfill the physical validity.

\indent (7) {\it Junction condition}: We have studied the junction condition for the
formation of thin shell between the interior and exterior spacetimes. Following the
condition of Darmois and Israel \cite{Darmois1927,Israel1966} we have studied the
surface energy density and surface pressure due to the formation of thin shell. The
variation of the surface energy density has been plotted in Figs. \ref{fig8} and 
it is observed that the similar behaviour can be obtained forsurface pressure. Thus, both
the parameters remain positive, which indicates that the thin shell satisfies the
weak and dominant energy conditions. From Eq. (\ref{eq70}) it can be claimed for
physically acceptable solution we must have $\frac{2M}{r}<1$ and $\rho_c(2\pi -Q)R^2<\frac{1}{2}$.

Unlike Einstein's general relativity there are different terms involving
the constant $Q$ in the expressions of different physical parameters of
the model. This is due to inclusion of the function of torsion scalar $\mathbb{T}$
as well as trace of the energy-momentum tensor $\mathcal{T}$ in the gravitational
Lagrangian of the corresponding action. This kind of modification has certainly
made the differences between the expressions in both the theories which can
be tested by doing a comparative study between this work and that of
Rahaman et al.~\cite{Rahaman2015} and Ghosh et al.~\cite{Ghosh2017} under
4-dimensional background.

\section*{ACKNOWLEDGMENTS}
SR is grateful to the Inter-University Centre for Astronomy and Astrophysics (IUCAA), Pune, 
India for providing the Visiting Associateship under which a part of this work was carried out.

\end{document}